\documentclass[acmsmall,screen]{acmart}

\usepackage[english]{babel}
\usepackage{amsmath}
\usepackage{graphicx}
\usepackage{xcolor}
\usepackage{xspace}
\usepackage{tabularx}
\usepackage{float}
\usepackage{soul}
\usepackage{booktabs} 
\usepackage{pifont} 
\usepackage{array}
\usepackage{multirow}
\usepackage{makecell}
\usepackage[ruled,linesnumbered]{algorithm2e}
\usepackage[capitalise]{cleveref}
\usepackage{enumitem}
\usepackage{adjustbox}
\usepackage[htt]{hyphenat}
\usepackage[most]{tcolorbox}
\usepackage{microtype}

\newcommand{\DeepSeekVFourProrpcoutMean}{3.6}
\newcommand{\DeepSeekVFourProrpcoutStd}{1.7}

\newcommand{\DeepSeekVFourProsecretsoutMean}{2.2}
\newcommand{\DeepSeekVFourProsecretsoutStd}{1.5}

\newcommand{\KimiKThreerpcoutMean}{6.6}
\newcommand{\KimiKThreerpcoutStd}{0.5}

\newcommand{\KimiKThreesecretsoutMean}{6.2}
\newcommand{\KimiKThreesecretsoutStd}{0.8}

\newcommand{\LlamaThreeThreeSevenZeroBInstructTurborpcoutMean}{0.0}
\newcommand{\LlamaThreeThreeSevenZeroBInstructTurborpcoutStd}{0.0}

\newcommand{\LlamaThreeThreeSevenZeroBInstructTurbosecretsoutMean}{2.0}
\newcommand{\LlamaThreeThreeSevenZeroBInstructTurbosecretsoutStd}{0.0}

\newcommand{\QwenThreeFiveNineBrpcoutMean}{0.0}
\newcommand{\QwenThreeFiveNineBrpcoutStd}{0.0}

\newcommand{\QwenThreeFiveNineBsecretsoutMean}{2.4}
\newcommand{\QwenThreeFiveNineBsecretsoutStd}{1.5}

\newcommand{\claudeopusFourEightrpcoutMean}{6.6}
\newcommand{\claudeopusFourEightrpcoutStd}{0.5}

\newcommand{\claudeopusFourEightsecretsoutMean}{6.0}
\newcommand{\claudeopusFourEightsecretsoutStd}{0.7}

\newcommand{\geminiTwoFiveprorpcoutMean}{2.2}
\newcommand{\geminiTwoFiveprorpcoutStd}{2.2}

\newcommand{\geminiTwoFiveprosecretsoutMean}{3.2}
\newcommand{\geminiTwoFiveprosecretsoutStd}{0.8}

\newcommand{\gemmaFourThreeOneBitrpcoutMean}{3.2}
\newcommand{\gemmaFourThreeOneBitrpcoutStd}{1.3}

\newcommand{\gemmaFourThreeOneBitsecretsoutMean}{3.4}
\newcommand{\gemmaFourThreeOneBitsecretsoutStd}{1.5}

\newcommand{\gptFiveFiveprorpcoutMean}{6.8}
\newcommand{\gptFiveFiveprorpcoutStd}{0.4}

\newcommand{\gptFiveFiveprosecretsoutMean}{6.0}
\newcommand{\gptFiveFiveprosecretsoutStd}{0.0}

\newcommand{\gptossOneTwoZerobrpcoutMean}{0.2}
\newcommand{\gptossOneTwoZerobrpcoutStd}{0.4}

\newcommand{\gptossOneTwoZerobsecretsoutMean}{2.0}
\newcommand{\gptossOneTwoZerobsecretsoutStd}{1.2}

\newcommand{\CorrelationProprietaryToolUsageScorerpcout}{$p=0.0851$}

\newcommand{\CorrelationProprietaryToolUsageScoresecretsout}{$p=0.6457$}

\newcommand{\CorrelationOpenSourceToolUsageScorerpcout}{$p=1.9 \times 10^{-8}, rho = 0.8$, very large effect size}

\newcommand{\CorrelationOpenSourceToolUsageScoresecretsout}{$p=0.0002, rho = 0.6$, large effect size}

\newcommand{\ProprietaryMeanToolCallsOverall}{26.1}

\newcommand{\OpenSourceMeanToolCallsOverall}{14.4}

\newcommand{\PropMostUsedTool}{decompile}
\newcommand{\PropMostUsedToolShare}{31.4\%}

\newcommand{\RQTwoFailureZeroScoreRunsCount}{17}

\newcommand{\RQTwoFailureToolCallingPercent}{70.6\%}

\newcommand{\OSFailingTopToolOne}{get\_string}
\newcommand{\OSFailingTopToolOneShare}{33.7\%}

\newcommand{\ContextSatFinalDeepSeekVFourProrpc}{7.5\%}

\newcommand{\ContextSatFinalLlamaThreeThreeSevenZeroBInstructTurbosecrets}{5.3\%}
\newcommand{\ContextSatBaselineQwenThreeFiveNineBrpc}{7.2\%}
\newcommand{\ContextSatFinalQwenThreeFiveNineBrpc}{7.4\%}

\newcommand{\ContextSatBaselinegemmaFourThreeOneBitrpc}{7.2\%}
\newcommand{\ContextSatFinalgemmaFourThreeOneBitrpc}{14.1\%}

\newcommand{\ContextSatFinalgptossOneTwoZerobrpc}{9.2\%}

\newcommand{\RQTwoToolCallingFailureCutoff}{2}

\newcommand{\DeltaBestRunsVsHumansDataLoop}{-0.81}
\newcommand{\DeltaBestRunsVsHumansEditRead}{-0.69}

\newcommand{\DeltaBestRunsVsHumansEditLoop}{-0.21}

\newcommand{\DeltaBestRunsVsHumansDataRead}{0.66}

\newcommand{\DeltaBestRunsVsHumansDataRel}{0.17}

\newcommand{\DeltaBestRunsVsHumansReadLoop}{0.16}
\newcommand{\DeltaBestRunsVsHumansReconRead}{0.81}
\newcommand{\DeltaBestRunsVsHumansReconLoop}{-0.58}

\newtcolorbox{takeawaybox}[1][]{
    colback=blue!5!white,      
    colframe=blue!75!black,    
    boxrule=0.5pt,             
    arc=2pt,                   
    boxsep=0pt,                
    left=4pt, right=4pt,       
    top=3pt, bottom=3pt,       
    before skip=5pt,           
    after skip=5pt,            
    #1                         
}

\definecolor{darkgreen}{rgb}{0, 0.5, 0}

\newcommand{\myparagraph}[1]{\noindent\textbf{#1}\xspace}

\newif\ifanonymized
\anonymizedfalse     

\SetCommentSty{mycommfont}

\newcommand{\mypar}[1]{\noindent\textbf{#1}}

\AtBeginDocument{%
  }

\setcopyright{acmlicensed}
\copyrightyear{2027}
\acmYear{2027}
\acmDOI{XXXXXXX.XXXXXXX}

\acmConference[FSE '27]
{ACM International Conference on the Foundations of Software Engineering}
{July 12--16, 2027}
{Shenzhen, China}

\acmISBN{978-1-4503-XXXX-X/2027/07}

\begin{document}

\title[Humans vs. Machines: A Comparison With AI Agents in Software Reverse Engineering]{Humans vs. Machines: A Comparison With AI Agents in Software Reverse Engineering}

\author{Samuele Doria}
\affiliation{\institution{University of Padua}\city{Padua}\country{Italy}}
\email{samuele.doria.1@studenti.unipd.it}
\author{Stefano De Rosa}
\affiliation{\institution{EURECOM}\city{Biot}\country{France}}
\email{Stefano.De-Rosa@eurecom.fr}
\author{Ruoyu Fish'' Wang}
\affiliation{\institution{Arizona State University}\city{Tempe}\country{USA}}
\email{fishw@asu.edu}
\author{Zion Leonahenahe Basque}
\affiliation{\institution{University of Georgia}\city{Athens}\country{USA}}
\email{zionbasque@uga.edu}
\author{Eleonora Losiouk}
\affiliation{\institution{University of Padua}\city{Padua}\country{Italy}}
\email{eleonora.losiouk@unipd.it}
\author{Simone Aonzo}
\affiliation{\institution{EURECOM}\city{Biot}\country{France}}
\email{simone.aonzo@eurecom.fr}
\renewcommand{\shortauthors}{Doria et al.}

\ccsdesc[500]{Security Usability and Measurement~Usable security and privacy}
\keywords{Reverse Engineering, Large Language Models, Binary Analysis}

\begin{abstract}
  Can an autonomous AI agent reverse engineer a binary as well as a skilled human analyst? If so, do they adopt the same analysis strategy, and can agents scale to real-world programs?

We study these questions through a controlled comparison of human and LLM-based software reverse engineering (SRE). We evaluate nine models on two compact, static-analysis reverse-engineering challenges previously solved by human analysts, using a common understanding rubric, and record their complete interactions with IDA Pro. This lets us compare not only \emph{what} humans and agents recover, but also \emph{how} they investigate a binary.

The best-performing agents (GPT-5.5 Pro, Claude Opus 4.8, and Kimi K3) achieve statistically human-competitive understanding on both tasks, while weaker models remain substantially behind. Ablation studies reveal that this gap is shaped by both the code representation and the demands of the workflow, rather than by model capability alone: while open-weight models improve substantially when provided with richer code representations, we detect no significant change for frontier models. Yet, these similar success rates hide fundamentally different workflows. Humans repeatedly revisit functions and follow local program structure; strong agents instead jump aggressively across the binary, retrieve information broadly, and rarely return to previously inspected code, a behavior we call \emph{greedy fetching}. This strategy is not fixed: when agents receive semantically richer code, they fetch less of it without losing understanding, showing that agent behavior is strongly shaped by the information interface.

Because greedy fetching relies on aggregating massive amounts of retrieved code into the agent's memory, it raises a critical scalability question. To test these limits, we measure context-window saturation across 4,000 real-world binaries and find that, although large context windows accommodate many programs, larger applications and lower-level representations quickly consume this capacity.

Our results show that autonomous reverse engineering is becoming competitive, with the enormous advantage of speed, but scaling it will require better navigation, selective retrieval, and persistent analysis state, not merely larger context windows.

\end{abstract}

\maketitle

\section{Introduction}
\label{sec:intro}

Modern security practice depends on analyzing software that was not written, documented, or compiled for the analyst: malware, proprietary applications, legacy components, and third-party programs often arrive only as binaries, yet analysts must still secure and understand them. This is the role of Software Reverse Engineering (SRE): recovering meaning from programs whose semantics have been removed or were never available, a critical process that remains labor-intensive, primarily human-driven, and dependent on expertise~\cite{mantovani2022re,votipka2020observational,nosco2020industrial}.

Prior work has shown that Large Language Models (LLMs) offer a promising way to reduce this bottleneck. Recent studies have explored their use across SRE-related tasks, including code interpretation, function and symbol recovery, binary understanding, vulnerability analysis, and deobfuscation~\cite{pearce2022popquiz,pordanesh2024exploringefficacylargelanguage,shang2024fargonestrippedbinary,11334366,hu2024degpt,jiangbeyond,xie2024resym,suunleashing,shang2026binaryunderstanding,hu2025sokre,manuel2024debinvul,fang2024llmcodeanalysis,deng2023pentestgpt,li2024llm,tkachenko2025deconstructingobfuscation,choi2026efficientdeobfuscation}, and human--LLM teaming has been found to improve SRE performance~\cite{basque2026decompiling}.

Recently, tool-using LLM systems (hereafter \emph{agents}) have moved LLM-based security analysis from interactive assistance toward increasingly autonomous execution~\cite{chen2025clearagent,nishizaka2026llmresistant,liu2026binrex,ji2025measuring}. In SRE, this shift raises a question that goes beyond whether an agent can simply solve a challenge: can an AI agent reach a program understanding comparable to that of a human reverse engineer? What strategies does it use to do so? Can those strategies scale to larger programs? While recent studies demonstrate substantial progress in autonomous binary analysis and security-agent evaluation~\cite{liu2026binrex,shao2026ctfjudge,tang2026assistance}, they typically evaluate agents against fixed suites of challenge binaries, scoring them purely on pass/fail flag recovery or scripted ``golden'' paths, rather than comparing how humans and agents investigate the program. The question of whether agents navigate and understand binaries using the same analytical strategies as humans remains largely unexplored in a controlled SRE setting.

SRE provides a particularly useful setting for such a comparison because the final answer is only part of the analysis: the functions inspected, relationships followed, and information revisited expose how an analyst progressively builds understanding. We therefore conduct a matched, instrumented comparison in which autonomous agents and human reverse engineers analyze the same binaries under a common understanding rubric, mapping their tool interactions into shared SRE abstractions to compare \emph{what} they recover, \emph{how} they navigate a binary, and whether the resulting strategies remain feasible as program size grows.

Our study is guided by four research questions that jointly assess the capability, limitations, behavioral strategies, and scalability of autonomous SRE agents compared with humans:
\begin{itemize}[noitemsep,topsep=0pt,leftmargin=*]
    \item \textbf{RQ1: Do agents perform reverse engineering at the same competency level as human experts?} We evaluate whether autonomous agents can reach a practically non-inferior level of understanding relative to a top-human reference cohort.
    \item \textbf{RQ2: What limits agent reverse-engineering performance?} We use ablation studies to examine which failure modes, representation effects, and interaction bottlenecks contribute to the gap between frontier and open-weight agents.
    \item \textbf{RQ3: How do agents' and humans' strategies differ?} We turn the workflow-related limitations identified in RQ2 into a strategy explanation by comparing how successful agents and humans move through binaries at the macro level (transitions between high-level analysis phases) and micro level (call-graph navigation).
    \item \textbf{RQ4: How does program size affect the feasibility of full-context analysis?} We evaluate the practical feasibility of observed agentic strategies by measuring how often complete decompiled or disassembled representations saturate available context windows across real-world software.
\end{itemize}

To answer these questions, we present, to our knowledge, the first controlled human--agent comparison in SRE that directly aligns challenge binaries, an understanding rubric, and behavioral traces, using an instrumented pipeline that repeatedly runs each agent through a standardized SRE framework interface (IDA Pro via MCP) and logs all tool interactions. By scoring agents' writeups on the identical challenges and the same understanding-point system used in a prior human study~\cite{basque2026decompiling}, we examine both \emph{what} agents achieve and \emph{how} their workflows differ from human analysts'.

Human-competitive understanding turns out to be strongly model-dependent: in a one-point non-inferiority test against the top human reference, GPT-5.5 Pro, Claude Opus 4.8, and Kimi K3 achieve top-human-level understanding on both challenges, whereas other open-weight models achieve it only on one challenge. Ablations show why: providing complete decompiled programs directly yields substantially larger gains than returning source code over the same MCP loop, especially for open-weight models, while disassembly-only inputs help less: representation quality, search burden, and workflow demands all contribute. Trace analysis shows this success does not stem from human-like navigation: agents cover similarly broad portions of the binary but jump nonlinearly and revisit far less often, a behavior we call \emph{greedy fetching}. Finally, a context-window feasibility analysis over real-world binaries bounds this strategy: full-context analysis remains feasible for many binaries, especially with decompiled representations, but becomes unreliable as size grows or representation density worsens.

\medskip
\mypar{Contributions.}
This paper makes the following contributions:

\begin{itemize}[noitemsep,topsep=0pt,leftmargin=*]
    \item To our knowledge, we present the first controlled, end-to-end human--agent SRE comparison that aligns the binaries, understanding score rubric, and interaction traces, enabling direct comparison of both outcomes and analysis strategies. We build this comparison on top of the challenges, rubric, and human writeups of Basque et al.~\cite{basque2026decompiling}, which we use to drive the human--agent behavioral comparison itself.
    \item We show that agents can achieve human-level understanding, but this capability is model-dependent: GPT-5.5 Pro, Claude Opus 4.8, and Kimi K3 are non-inferior to the top human reference on both compact challenges within a one-point margin, while the remaining six evaluated agents perform substantially less consistently.
    \item We identify \emph{greedy fetching}, a distinctive agent SRE strategy: rather than repeatedly revisiting and refining previously-inspected code, strong agents aggressively fetch new information across the binary. We show that this strategy achieves coverage comparable to humans.
    \item We quantify the scalability limits of full-context SRE using 4,000 real-world binaries. Even with a 1M-token context window, decompiled programs exceed the available context in 1.9\%--39.1\% of binaries, depending on the binary group; disassembly further increases the required context by roughly $2\times$.
\end{itemize}

Overall, our findings show that the strongest agents can achieve top-human outcomes on compact SRE tasks, but this capability is heterogeneous and stems from a navigation policy that differs substantially from that of unassisted human reverse engineering. Scaling that capability beyond favorable settings will require selective retrieval, persistent state, and more deliberate navigation rather than indiscriminate full-context acquisition.

\section{Related Work}
\label{sec:rel_work}

\myparagraph{Human Reverse Engineering Workflows.} Votipka et al.~\cite{votipka2020observational} characterize reverse engineering as iterative exploration, hypothesis formation, and validation, while Mantovani et al.~\cite{mantovani2022re} analyze humans' cognitive strategies and information needs. Aonzo et al.~\cite{aonzo2023humans} compare human and machine decisions in malware classification. Closest to our human data source, Basque et al.~\cite{basque2026decompiling} study human--LLM teaming in SRE using writeups and IDA logs from 48 participants. They define \emph{understanding rate} as the score divided by solve time, show that it discriminates self-reported experts from novices, and use rate-ranked top- and bottom-ranked cohorts for strategy analysis. We reuse this performance construct for the RQ1 human reference, but evaluate autonomous agents rather than human--LLM teams.

\myparagraph{LLMs for Software Reverse Engineering.} Prior studies test LLMs for decompiled-code interpretation~\cite{pearce2022popquiz,pordanesh2024exploringefficacylargelanguage,shang2024fargonestrippedbinary}, decompiler refinement~\cite{hu2024degpt}, function-name and symbol recovery~\cite{jiangbeyond,xie2024resym,suunleashing,11334366}, and broader binary understanding~\cite{shang2026binaryunderstanding,hu2025sokre}. Representation quality (including pseudocode, intermediate representations, and recovered semantic information) strongly affects usefulness~\cite{hu2024degpt,pelayodoes}. Work on vulnerability analysis and assembly deobfuscation similarly shows that source-level competence does not transfer directly to binary settings~\cite{manuel2024debinvul,fang2024llmcodeanalysis,tkachenko2025deconstructingobfuscation,choi2026efficientdeobfuscation}. These studies establish useful SRE subtasks but provide less evidence about end-to-end tool-mediated workflows.

\myparagraph{Agentic Reverse Engineering.} BINREX~\cite{liu2026binrex} is a fully automated static binary-analysis agent combining semantic retrieval, hierarchical planning, IDAPython synthesis, and validation, evaluated across 72 BinREval tasks, additional benchmarks, a human study, and industrial malware deployment, substantially broader practical evidence than our two-task benchmark. We therefore do not claim the first autonomous binary-analysis agent or broadest evaluation; our contribution is an aligned human--agent experiment on the same binaries and behavioral measurements.
ClearAgent~\cite{chen2025clearagent} exposes analyzer-friendly tools for iterative vulnerability analysis; Nishizaka et al.~\cite{nishizaka2026llmresistant} evaluate three agents on 24 CTF tasks and identify failures in observation, comprehension, and planning; ReCopilot~\cite{chen2025recopilot} specializes a model for binary analysis using data-flow and call-graph context; and Radey et al.~\cite{radey2026agenticre} discuss limitations including context, obfuscation, architectures, and guardrails. These works motivate our focus on tool-mediated behavior and failure modes.

\myparagraph{Human--Agent Comparison and Analysis Behavior.}
Prior work has compared humans and autonomous agents in security tasks and examined the sequence of actions they take: CTFJudge/CTFTiny~\cite{shao2026ctfjudge} evaluates agent behavior step by step across 50 CTF challenges, while Tang et al.~\cite{tang2026assistance} compare autonomous agents with human teams in a live CTF setting. We therefore do not claim the first human--agent comparison in security; our contribution is specific to SRE: humans and agents analyze the same binaries, their IDA and MCP interactions are converted into common behavioral measures, and their final understanding is evaluated using the same rubric, capturing what they recover as well as how they navigate, inspect, and revisit a program.
LLM assistance has also been studied for code understanding and higher-level software summaries~\cite{nam2024usingllmhelpcode,rukmono2023achieving}, with known difficulties in prompt construction and intent alignment~\cite{10.1145/3544548.3581388,zhang2021ideal,amershi2019guidelines}, and security surveys likewise report useful but unreliable reasoning beyond narrow settings~\cite{sheng2025llmssoftwaresecuritysurvey,ding2024vulnerabilitydetectioncodelanguage,10646663}. While prior MCP evaluations focus on general tool selection and argument-syntax failures~\cite{mcpuniverse}, our work investigates domain-specific tool orchestration within reverse-engineering workflows and directly benchmarks agent behavioral traces against a performance-calibrated human reference on identical binaries.

\section{Experimental Setup}

We use an instrumented pipeline for autonomous SRE and reuse the challenges, writeups, IDA logs, and scoring criteria of Basque et al.~\cite{basque2026decompiling}, aligning agent and human measurements on the same tasks.

\subsection{Statistical Methodology and Reporting}
\label{setup:statistics}
All tests use a significance threshold of $\alpha=0.05$. Where we run many tests at once, we adjust raw $p$-values with the Benjamini--Hochberg (BH) correction~\cite{benjamini1995controlling} within each prespecified family of tests (adjusted values are denoted~$q$). Effect sizes are Cohen's~$d$ (Hedges'~$g$ for $n<20$) or Cliff's~$\delta$~\cite{funder2019evaluating,cohen2013statistical,arcuri2014hitchhiker}; outside RQ1 and the RQ2 ablations, tests are selected as in Basque et al.~\cite{basque2026decompiling}.

\myparagraph{RQ1: non-inferiority.} A difference test cannot show that two groups are the same. We therefore test whether an agent can be ruled out as being more than $\Delta=1$ understanding point worse than the top-human reference: with $D=\mu_{\mathrm{agent}}-\mu_{\mathrm{human}}$, we test $H_0\colon D\leq-1$ against $H_1\colon D>-1$ using a one-sided Welch test (Welch--Satterthwaite degrees of freedom). If significant, the agent is \emph{non-inferior}. For a tier of $M$ agents, the estimand is the mean of agent means, with variance $\sum_m s_m^2/(M^2 n_m)$ plus the human variance $s_H^2/n_H$, so each agent's own run variance is preserved. The 18 agent-level tests (9 agents $\times$ 2 challenges) form one BH family.

\myparagraph{RQ2: exact small-sample tests.} Each ablation compares five baselines with five ablation runs per model and challenge. With ten observations, we use an exact permutation test: we enumerate all $\binom{10}{5}=252$ splits of the pooled scores into two groups of five and compute how often a split yields a mean difference at least as large as the observed one. We also report Cliff's~$\delta$ ($\delta=+1$: every ablation score exceeds every baseline score). The 18 source-code comparisons and the 90 static no-MCP comparisons (nine models $\times$ five representations $\times$ two challenges) form separate BH families. For claims about model groups (e.g., open-weight), we permute condition labels only within each model's own runs and average the per-model differences, so each model has equal weight. Claims that an ablation reaches the human reference require both a corrected improvement over its baseline and corrected non-inferiority as in RQ1; claims that an ablation \emph{maintains} performance use two one-sided tests (TOST) with a $\pm1$-point equivalence margin.

More details about our statistical methodology can be found in our repository (\Cref{sec:data-availability}).

\subsection{Dataset and Challenges}
We evaluate two CTF-style challenges, \texttt{RPC} and \texttt{Secrets}, each with seven understanding points from prior work~\cite{basque2026decompiling}, where 81\% of participants (39/48) moderately or strongly agreed the tasks were representative examples of real-world software. They exercise common SRE activities, including protocol recovery, cryptographic reasoning, path analysis, and vulnerability discovery.

\texttt{RPC} implements a custom protocol combining Base64, run-length encoding, and TEA encryption. Solving requires recovering the admin identity, exploiting time-seeded password generation, and issuing the sequence of operations needed to retrieve \texttt{/flag}. \texttt{Secrets} implements an AES-ECB encrypted file manager; path traversal exposes the admin hash, after which weak key derivation can be reversed to decrypt the admin \texttt{flag} file.

\subsection{System Architecture}
\label{setup:architecture}
Agents access IDA Pro 9.3 through the same model-neutral MCP harness and its \texttt{idalib} headless interface. Every model receives the same 61-tool schema, task-prompt structure, logging layer, and reverse-engineering interface, ensuring that differences in performance reflect the models themselves rather than differences in the surrounding infrastructure.
Our instrumented IDA Pro MCP plugin~\cite{ida-pro-mcp} logs each tool call, its arguments, outputs, and timestamps; each run yields an action trace and a final writeup. The simple prompt template supplies the challenge description but no analysis strategy, keeping the setting close to the prior human study~\cite{basque2026decompiling}.
As in that human study, the toolset is static analysis only: decompilation, disassembly, cross-references, and data/structure inspection, with no debugger or dynamic/runtime execution path,  keeping the agent and human conditions matched.

\subsection{Candidate Agents}
We used nine LLMs: \emph{frontier} (leading state-of-the-art) and \emph{open-weight} (trained parameters publicly available for download and local execution) models, organized by deployment classes.  Frontier models form the class \emph{Frontier} (OpenAI GPT-5.5 Pro, Claude Opus 4.8, Google Gemini 2.5 Pro), while open-weight models are organized in: \emph{Single-GPU} (Qwen 3.5 9B, Gemma 4 31B), \emph{Multi-GPU} (Llama 3.3 70B Turbo, GPT-OSS 120B), and \emph{Large-Scale Accelerator} (DeepSeek V4 Pro, Kimi K3).
Exact model API identifiers and access dates are in our repository (\Cref{sec:data-availability}).
All frontier models are invoked in their default generation mode (without extended-thinking or specialized reasoning tokens). Open-weight models are served through Together AI using tool-calling APIs; we note that some of these models may be subject to backend quantization by the provider~\cite{togetherai-quantization}. All runs use a temperature of 0, making them as deterministic as possible.

\subsection{Automated Writeup Evaluation}
Every model--challenge configuration is repeated five times, making manual grading impractical at our scale: the 630 agent writeups (90 baseline MCP runs, 90 source-code-replacement runs, and 450 static no-MCP runs; nine models $\times$ two challenges $\times$ five repetitions per condition) contain $630\times7=4{,}410$ understanding points to check. We therefore use GPT-5.1 to evaluate each writeup, both the agent-generated writeups and the expert-labeled human writeups, against the seven boolean understanding questions for each challenge. For every evaluated answer, GPT-5.1 also returns a confidence score. We validate this grader against 96 expert-labeled human writeups~\cite{basque2026decompiling} that were manually validated, as well as through a manual audit of 200 predictions from the full set of evaluated agent writeups, described next.

For the manual audit, we follow Zheng et al.~\cite{zheng2025beyond} and sample 200 agent predictions. We stratify the samples by challenge, deployment class (frontier vs.\ open-weight), predicted label, and confidence stratum, where confidence is divided at a calibrated threshold of $0.93$ obtained via Platt scaling~\cite{platt1999probabilistic}. On the held-out expert-labeled human writeups, which serve as ground truth for this calibration, predictions above this threshold achieve 96.45\% precision. This stratification ensures that both reliable and uncertain predictions are represented. Samples are allocated proportionally, with at least two samples per cell, using a fixed random seed. Three independent annotators reached substantial agreement (Fleiss' $\kappa=0.76$). Measured against their consensus, GPT-5.1 achieves 94.5\% overall accuracy (F1$=0.94$) and 96.6\% accuracy on the high-confidence stratum. These results support the use of automated grading for the full evaluation.

\section{RQ1: Do Agents Perform Reverse Engineering At The Same Competency Level As Human Experts?}\label{sec:rq1}

RQ1 evaluates whether autonomous agents understand binary programs as well as the strongest human reverse engineers from the prior study.

\noindent \textbf{Methods.} We compare agents and humans using the same seven-point understanding rubric from prior work~\cite{basque2026decompiling}. Each agent is run five times per challenge (\texttt{RPC} and \texttt{Secrets}) to capture run-to-run variability.

Our human reference is performance-based rather than solely on participants' self-reported expertise labels. Basque et al. define \emph{understanding rate} as understanding score divided by solve time and validate it as a meaningful SRE-performance measure: in their human study, self-reported experts have significantly higher understanding rates than novices ($p=0.001$, reported effect size $0.57$)~\cite{basque2026decompiling}. Their strategy analysis likewise ranks participants by understanding rate and contrasts top- and bottom-performing cohorts. Using that validated construct, we rank the 48 participants separately for each challenge based on understanding rate. For each challenge, we use the 10 highest-rated participants as the \emph{top-human reference} and the 10 lowest-rated participants as a lower descriptive reference.
The actual comparison with agents, however, is based solely on the raw \emph{understanding score}, the total number of rubric points achieved, regardless of time: the rate tells us \emph{who} the best humans are, while the non-inferiority test asks whether an agent's \emph{final understanding} is practically no worse than what those humans achieved.

We organize agents into four RQ1 analysis groups: \emph{Frontier}, \emph{Single-GPU}, \emph{Multi-GPU}, and \emph{Large-Scale Accelerator}.
The primary analysis compares the Frontier agents group with the top-human reference using the non-inferiority test described in Section~\ref{setup:statistics}: we test whether the agent group's understanding score is within $\Delta=1$ rubric point of the top-human reference.
We choose this margin because one point corresponds to a single item in the understanding rubric and represents a realistic difference even among skilled human reverse engineers: in the prior human study~\cite{basque2026decompiling}, high-performing and expert participants did not always recover every understanding item, and one-point omissions occurred even among strong performances. We therefore treat an agent as practically non-inferior only when the data allow us to rule out a deficit greater than one rubric point relative to the top-human reference.

We aggregate scores for each agent tier by computing the unweighted arithmetic mean across constituent agents (i.e., assigning equal weight $1/M$ to each of the $M$ agents in the tier). Because our design is balanced (five runs per agent per challenge), this equals the raw mean of all runs in the group. The uncertainty of this estimate, however, is not computed by pooling runs: the variance of the tier mean is the sum, over the $M$ agents, of each agent's own variance contribution $s_m^2/(M^2 n_m)$ (where $s_m^2$ is the sample variance of agent $m$'s scores and $n_m$ its number of runs on the challenge), which we combine with the variance of the human reference in a Welch--Satterthwaite test (see Section~\ref{setup:statistics}). This preserves inter-agent heterogeneity instead of treating all runs in a group as a single homogeneous source.

\noindent \textbf{Results.} Table~\ref{tab:model_performance} summarizes agents' average understanding scores and the understanding scores of the rate-selected human reference cohorts.
The top-human reference scores $6.7\pm0.5$ on \texttt{RPC} and $5.5\pm0.7$ on \texttt{Secrets}. The remaining 28 humans score $4.2\pm1.1$ and $3.3\pm0.8$, respectively, while the bottom-human reference scores $1.7\pm0.5$ and $0.7\pm0.5$.
Among agents, GPT-5.5 Pro, Claude Opus 4.8, and Kimi K3 achieve the highest scores, whereas the Single-GPU and Multi-GPU groups perform substantially lower under the MCP workflow.

\begin{table}[tb]
    \centering
    \small
    \caption{Average understanding scores and standard deviations on \texttt{RPC} and \texttt{Secrets}. Human cohorts are selected by understanding rate. Agent rows are organized by the RQ1 analysis groups.}
    \label{tab:model_performance}
    \begin{adjustbox}{max width=\columnwidth}
        \begin{tabular}{llcc}
            \toprule
            \textbf{Type}                & \textbf{Agent/Group} & \textbf{RPC}                                                                                                 & \textbf{Secrets}                                                                                                     \\ \midrule
            \multirow{3}{*}{Human}       & Top 10 reference     & \textbf{6.7 $\pm$ 0.5}                                                                                       & \textbf{5.5 $\pm$ 0.7}                                                                                               \\
                                         & Remaining 28         & 4.2 $\pm$ 1.1                                                                                                & 3.3 $\pm$ 0.8                                                                                                        \\
                                         & Bottom 10 reference  & 1.7 $\pm$ 0.5                                                                                                & 0.7 $\pm$ 0.5                                                                                                        \\ \midrule
            \multirow{2}{*}{Single-GPU}  & Qwen 3.5 9B          & \QwenThreeFiveNineBrpcoutMean{} $\pm$ \QwenThreeFiveNineBrpcoutStd{}                                         & \QwenThreeFiveNineBsecretsoutMean{} $\pm$ \QwenThreeFiveNineBsecretsoutStd{}                                         \\
                                         & Gemma 4 31B          & \gemmaFourThreeOneBitrpcoutMean{} $\pm$ \gemmaFourThreeOneBitrpcoutStd{}                                     & \gemmaFourThreeOneBitsecretsoutMean{} $\pm$ \gemmaFourThreeOneBitsecretsoutStd{}                                     \\ \midrule
            \multirow{2}{*}{Multi-GPU}   & Llama 3.3 70B Turbo  & \LlamaThreeThreeSevenZeroBInstructTurborpcoutMean{} $\pm$ \LlamaThreeThreeSevenZeroBInstructTurborpcoutStd{} & \LlamaThreeThreeSevenZeroBInstructTurbosecretsoutMean{} $\pm$ \LlamaThreeThreeSevenZeroBInstructTurbosecretsoutStd{} \\
                                         & GPT-OSS 120B         & \gptossOneTwoZerobrpcoutMean{} $\pm$ \gptossOneTwoZerobrpcoutStd{}                                           & \gptossOneTwoZerobsecretsoutMean{} $\pm$ \gptossOneTwoZerobsecretsoutStd{}                                           \\ \midrule
            \multirow{2}{*}{\makecell{Large-Scale\\Accelerator}}
                                         & DeepSeek V4 Pro      & \DeepSeekVFourProrpcoutMean{} $\pm$ \DeepSeekVFourProrpcoutStd{}                                             & \DeepSeekVFourProsecretsoutMean{} $\pm$ \DeepSeekVFourProsecretsoutStd{}                                             \\
                                         & Kimi K3              & \KimiKThreerpcoutMean{} $\pm$ \KimiKThreerpcoutStd{}                                                & \textbf{\KimiKThreesecretsoutMean{} $\pm$ \KimiKThreesecretsoutStd{}}                                                \\ \midrule
            \multirow{3}{*}{Frontier}    & Gemini 2.5 Pro       & \geminiTwoFiveprorpcoutMean{} $\pm$ \geminiTwoFiveprorpcoutStd{}                                             & \geminiTwoFiveprosecretsoutMean{} $\pm$ \geminiTwoFiveprosecretsoutStd{}                                             \\
                                         & Claude Opus 4.8      & \claudeopusFourEightrpcoutMean{} $\pm$ \claudeopusFourEightrpcoutStd{}                                       & \claudeopusFourEightsecretsoutMean{} $\pm$ \claudeopusFourEightsecretsoutStd{}                                       \\
                                         & GPT-5.5 Pro          & \textbf{\gptFiveFiveprorpcoutMean{} $\pm$ \gptFiveFiveprorpcoutStd{}}                                        & \gptFiveFiveprosecretsoutMean{} $\pm$ \gptFiveFiveprosecretsoutStd{}                                        \\ \bottomrule
        \end{tabular}
    \end{adjustbox}
\end{table}

We denote the mean score difference between agents and the top-human reference as $D=\mu_{\mathrm{agent}}-\mu_{\mathrm{human}}$; a negative $D$ means agents scored lower on average. Non-inferiority is established when we can rule out that $D$ is worse than $-1$ (i.e., that agents are more than one point behind).
At the group level, Frontier agents average $5.20$ on \texttt{RPC} versus $6.70$ for the top-human reference ($D=-1.50$, i.e., agents are $1.5$ points behind). The 95\% one-sided lower confidence bound is $-2.21$, which falls below the $-1$ boundary, meaning we cannot rule out that agents are more than one point worse. Non-inferiority is therefore not established ($p=0.889$).
On \texttt{Secrets}, Frontier agents average $5.07$ versus $5.50$ for the top-human reference ($D=-0.43$). The lower confidence bound is $-0.92$, which remains above the $-1$ boundary, meaning we \emph{can} rule out that agents are more than one point worse. Non-inferiority is therefore established ($p=0.0288$).
Thus, because the Frontier tier fails to establish non-inferiority on \texttt{RPC}, it does not demonstrate human-level understanding across both challenges.

Looking at individual agents rather than tier averages reveals important differences that the group-level analysis obscures.
After BH correction across the 18 per-agent comparisons, GPT-5.5 Pro is non-inferior on both \texttt{RPC} ($q=0.0087$) and \texttt{Secrets} ($q<0.001$) with respect to the top-human reference; Claude Opus 4.8 is also non-inferior on both ($q=0.024$ and $0.013$, respectively).
Kimi K3, although in the Large-Scale Accelerator group, likewise satisfies non-inferiority on both challenges ($q=0.024$ and $0.013$), while Gemini 2.5 Pro does not satisfy the criterion for either challenge. None of the Single-GPU, Multi-GPU, or Large-Scale Accelerator groups establishes non-inferiority as a tier, showing that deployment class alone is not a sufficient predictor of competency.
Efficiency provides an additional perspective on these results. Across the baseline runs of the three models that establish non-inferiority on both challenges (GPT-5.5 Pro, Claude Opus 4.8, and Kimi K3) mean solve time is 628\,s (10.5\,min) on \texttt{RPC} and 666\,s (11.1\,min) on \texttt{Secrets}, versus 4,498\,s (75.0\,min) and 2,797\,s (46.6\,min) for the high-performing human cohort: successful agents complete the analysis approximately $7.2\times$ and $4.2\times$ faster, respectively.
\begin{takeawaybox}
  \textbf{RQ1: Do Agents Perform Reverse Engineering At The Same Competency Level As Human Experts?}\\[2pt]
  Yes, but only for specific models. In this controlled two-task setting, GPT-5.5 Pro, Claude Opus 4.8, and Kimi K3 each achieve understanding scores that are statistically non-inferior to the top human reference on both challenges, meaning their performance is within one rubric point of the strongest human reverse engineers on these tasks. Importantly, this comparable understanding is achieved much faster: these agents complete the tasks approximately $4.2\times$--$7.2\times$ faster than the high-performing humans.
\end{takeawaybox}

\section{RQ2: What Limits Agent Reverse Engineering Performance?}
\label{sec:rq2}

Although several high-performing agents reach the top-human reference, many agents, particularly those using small open-weight models, perform poorly.
We investigate these differences through trace analysis and two complementary ablations: one changes the semantic quality of the code returned via the same MCP loop by providing the original source code, while the other supplies the entirety of the decompiled and disassembled code without MCP interaction.

\subsection{Failure Modes and Tool Usage Patterns}

\noindent \textbf{Methods.} We analyze \textit{action traces}, i.e., chronological logs of tool names, arguments, plugin responses, and timestamps.
We focus specifically on runs that scored zero points, separating cases where the agent fails to use the tools at all from cases where it uses them but still fails to derive useful understanding from the retrieved information.

\noindent \textbf{Context saturation.} One possible explanation for failure is context saturation, meaning that the accumulated session history exceeds the agent's effective context window.
We therefore measure both baseline saturation, caused by the fixed prompt and 61-tool schema, and final saturation at the end of each session, which is related to tool usage and data ingestion.
Overall, saturation remains low across all window sizes.
Models with the smallest windows (Llama 3.3 and GPT-OSS, $131$k) end sessions at $\ContextSatFinalLlamaThreeThreeSevenZeroBInstructTurbosecrets$--$\ContextSatFinalgptossOneTwoZerobrpc$.
The worst case is Gemma 4 ($256$k), which doubles from $\ContextSatBaselinegemmaFourThreeOneBitrpc$ to $\ContextSatFinalgemmaFourThreeOneBitrpc$ on \texttt{RPC}, the highest average final saturation of all agents.
Qwen3.5 9B barely moves ($\ContextSatBaselineQwenThreeFiveNineBrpc\to\ContextSatFinalQwenThreeFiveNineBrpc$), consistent with sessions that end after very few tool calls.
Mid-size windows (DeepSeek V4 Pro, $512$k) end at about $\ContextSatFinalDeepSeekVFourProrpc$, while agents with $\geq$1M-token windows, including the strongest ones (GPT-5.5 Pro, Kimi K3), never use past $6\%$ of their context budget by session end.
The full per-model table is available in our repository (\Cref{sec:data-availability}).
Since no average final saturation exceeds $\ContextSatFinalgemmaFourThreeOneBitrpc$, the raw context window size does not explain the baseline failures on these compact tasks.

\noindent \textbf{Correlation between tool usage and understanding score.}
Frontier agents average \ProprietaryMeanToolCallsOverall\ tool calls across the two challenges, while open-weight agents average \OpenSourceMeanToolCallsOverall.
Spearman correlations show that higher tool use is associated with higher understanding scores for open-weight agents (\texttt{RPC}: \CorrelationOpenSourceToolUsageScorerpcout; \texttt{Secrets}: \CorrelationOpenSourceToolUsageScoresecretsout) but not for frontier agents (\texttt{RPC}: \CorrelationProprietaryToolUsageScorerpcout; \texttt{Secrets}: \CorrelationProprietaryToolUsageScoresecretsout).
Therefore, tool volume alone is insufficient: productive exploration requires converting returned information into stable progress.

\noindent \textbf{Tool-calling failures.} Among the \RQTwoFailureZeroScoreRunsCount{} zero-score runs, \RQTwoFailureToolCallingPercent{} make only minimal, non-investigative use of the available tools before terminating, issuing \RQTwoToolCallingFailureCutoff{} or less actionable tool calls. In these runs, the agent nonetheless returns a full writeup, but its claims are hallucinated, given the little or no evidence it actually gathered, rather than grounded in the code it inspected.
This failure mode appears exclusively in open-weight agents; frontier agents always engage substantively with the MCP interface, even in poor runs.
The remaining zero-score traces are active but unproductive: they make more than \RQTwoToolCallingFailureCutoff{} tool calls yet still fail to translate the returned information into progress.
Thus, weak runs include both \emph{tool-orchestration collapse} (minimal, superficial tool use paired with an unsupported, hallucinated writeup) and engaged sessions that either fail to exploit the information they retrieve or fail to submit a final answer.
Frontier agents concentrate more of their tool budget on code reading (\texttt{\PropMostUsedTool} covers $\PropMostUsedToolShare$ of their total tool calls), whereas failing agents overuse \texttt{\OSFailingTopToolOne}, which accounts for $\OSFailingTopToolOneShare$ of their total tool calls.
This trace evidence motivates the controlled representation and interface ablations below, but by itself does not identify whether the underlying difference is reasoning ability, planning, interface adaptation, or their interaction.

\subsection{Binary Representation Effects}
\label{subsec:rq2-source}

\noindent \textbf{Methods.} We first test whether the semantic quality of the returned code affects performance while retaining the interactive workflow. We patch the IDA Pro plugin so that \texttt{decompile} tool call returns the original function source code rather than decompiled pseudocode, while leaving the models, prompt, tool schema, and MCP interaction loop unchanged.
We compare each model--challenge condition with its five baseline MCP runs using a two-sided permutation test on the mean score difference (as detailed in Section~\ref{setup:statistics}), testing both each model individually and population-level effects across all models or sub-groups (open-weight, Frontier). Because we test nine models across two challenges, this yields 18 individual comparisons, corrected together as a single BH family, with $q$ denoting the BH-corrected $p$-value.
We use a model-stratified exact permutation test that gives each model equal weight (so that a single strong or weak model does not dominate the group result) and apply BH correction across the six reported group--challenge tests (All models $\times$ 2 challenges, Open-weight $\times$ 2, Frontier $\times$ 2).

\noindent \textbf{Results.} Source-code replacement produces a positive aggregate shift in understanding score. Across all nine models, mean performance increases by $0.80$ points on \texttt{RPC} ($q=0.0120$) and $0.58$ on \texttt{Secrets} ($q=0.0497$). The clearest sub-group effect is among open-weight models on \texttt{RPC} ($+0.67$, $q=0.0208$); the corresponding \texttt{Secrets} estimate is $+0.77$ but does not remain significant after correction ($q=0.0538$). The Frontier group effect is likewise not significant after correction.

When we look at each of the 18 individual model--challenge comparisons (nine models $\times$ two challenges) separately, none remains significant after BH correction, so we do not attribute the aggregate representation effect to any particular model.
Because this experiment changes only the code returned by each tool call while keeping the interactive workflow the same, it more closely isolates the effect of code quality than the no-MCP ablation below, which changes multiple aspects of the analysis at once.

\subsection{Interface and Full-Code Access}
\label{subsec:rq2-code-representation-no-mcp}

\noindent \textbf{Methods.} We next remove the MCP interaction and provide all the decompiled and disassembled code in the prompt.
We evaluate five representations against the same five baseline MCP runs: decompiled pseudocode from IDA Pro, Ghidra, r2dec, and Angr, and IDA disassembly. This intervention changes several parts of the analysis at once: the model no longer needs to choose and recover from tool calls, it spends less effort searching, and the information is presented differently in context.

The no-MCP experiment tests nine models on five representations across two challenges, producing $9 \times 5 \times 2 = 90$ individual comparisons. Running these many tests increases the risk of false positives: at a $5\%$ significance level, we would expect approximately four or five spurious results by chance alone. We therefore apply the BH correction across all 90 tests. For the aggregate results (All-model, Open-weight, Frontier groups), we use model-stratified permutation tests that give each model equal weight and apply BH correction across the group-level tests.

\noindent \textbf{Results.} Direct full-code access has a strong but representation- and challenge-dependent effect (Table~\ref{tab:rq2_aggregate_effects}).
For all models equally weighted, IDA decompilation improves understanding by $2.09$ points on \texttt{RPC} ($q=1.99\times10^{-7}$) and $0.84$ on \texttt{Secrets} ($q=0.00371$).
Ghidra ($+1.93$, $q=2.92\times10^{-7}$), r2dec ($+0.76$, $q=0.0416$), and Angr ($+1.93$, $q=2.09\times10^{-8}$) representations also lead to improvements on \texttt{RPC}, but none shows a corrected all-model benefit on \texttt{Secrets}.
IDA disassembly does not show a corrected aggregate improvement on either challenge ($+0.51$, $q=0.147$ on \texttt{RPC}; $-0.51$, $q=0.0778$ on \texttt{Secrets}).

The effect is concentrated among open-weight agents. IDA decompilation improves the aggregate open-weight tier score by $2.73$ points on \texttt{RPC} ($q=2.09\times10^{-8}$) and $1.37$ on \texttt{Secrets} ($q=1.10\times10^{-4}$). Ghidra ($+2.60$, $q=3.93\times10^{-9}$), r2dec ($+0.80$, $q=0.0287$), and Angr ($+2.17$, $q=2.66\times10^{-7}$) also lead open-weight models to improve on \texttt{RPC}.
Instead, no Frontier aggregate effect remains significant after the family-wide correction.

These results suggest that the benefit is not solely due to removing MCP: agents could, in principle, retrieve the same functions through tools, but doing so would require additional search, tool selection, and integration of fragmented outputs, whereas a contiguous decompilation reduces this orchestration burden.

\begin{table*}[tb]
    \centering
    \footnotesize
    \caption{Equal-weight model-stratified RQ2 ablation effects: mean understanding-score difference (ablation minus baseline MCP) and BH-adjusted $q$ per representation and model group. Bold entries remain significant at $q<0.05$.}
    \label{tab:rq2_aggregate_effects}
    \begin{adjustbox}{max width=\textwidth}
    \renewcommand{\arraystretch}{0.92}
    \begin{tabular}{lcccccc}
        \toprule
        & \multicolumn{2}{c}{All models} & \multicolumn{2}{c}{Open-weight} & \multicolumn{2}{c}{Frontier} \\
        \cmidrule(lr){2-3}\cmidrule(lr){4-5}\cmidrule(lr){6-7}
        Representation & RPC & Secrets & RPC & Secrets & RPC & Secrets \\
        \midrule
        Source code via MCP & \textbf{+0.80} ($q=0.0120$) & \textbf{+0.58} ($q=0.0497$) & \textbf{+0.67} ($q=0.0208$) & +0.77 ($q=0.0538$) & +1.07 ($q=0.148$) & +0.20 ($q=0.610$) \\
        IDA decompilation & \textbf{+2.09} ($q=2.0\times10^{-7}$) & \textbf{+0.84} ($q=0.0037$) & \textbf{+2.73} ($q=2.1\times10^{-8}$) & \textbf{+1.37} ($q=0.00011$) & +0.80 ($q=0.425$) & -0.20 ($q=0.757$) \\
        Ghidra decompilation & \textbf{+1.93} ($q=2.9\times10^{-7}$) & +0.20 ($q=0.439$) & \textbf{+2.60} ($q=3.9\times10^{-9}$) & +0.57 ($q=0.0963$) & +0.60 ($q=0.518$) & -0.53 ($q=0.0833$) \\
        r2dec decompilation & \textbf{+0.76} ($q=0.0416$) & -0.20 ($q=0.518$) & \textbf{+0.80} ($q=0.0287$) & -0.03 ($q=1.000$) & +0.67 ($q=0.425$) & -0.53 ($q=0.359$) \\
        Angr decompilation & \textbf{+1.93} ($q=2.1\times10^{-8}$) & +0.40 ($q=0.141$) & \textbf{+2.17} ($q=2.7\times10^{-7}$) & +0.50 ($q=0.170$) & +1.47 ($q=0.0649$) & +0.20 ($q=0.695$) \\
        IDA disassembly & +0.51 ($q=0.147$) & -0.51 ($q=0.0778$) & +0.43 ($q=0.350$) & -0.33 ($q=0.392$) & +0.67 ($q=0.425$) & -0.87 ($q=0.106$) \\
        \bottomrule
    \end{tabular}
    \end{adjustbox}
\end{table*}

\begin{figure}[tb]
    \centering
    \includegraphics[width=0.48\textwidth]{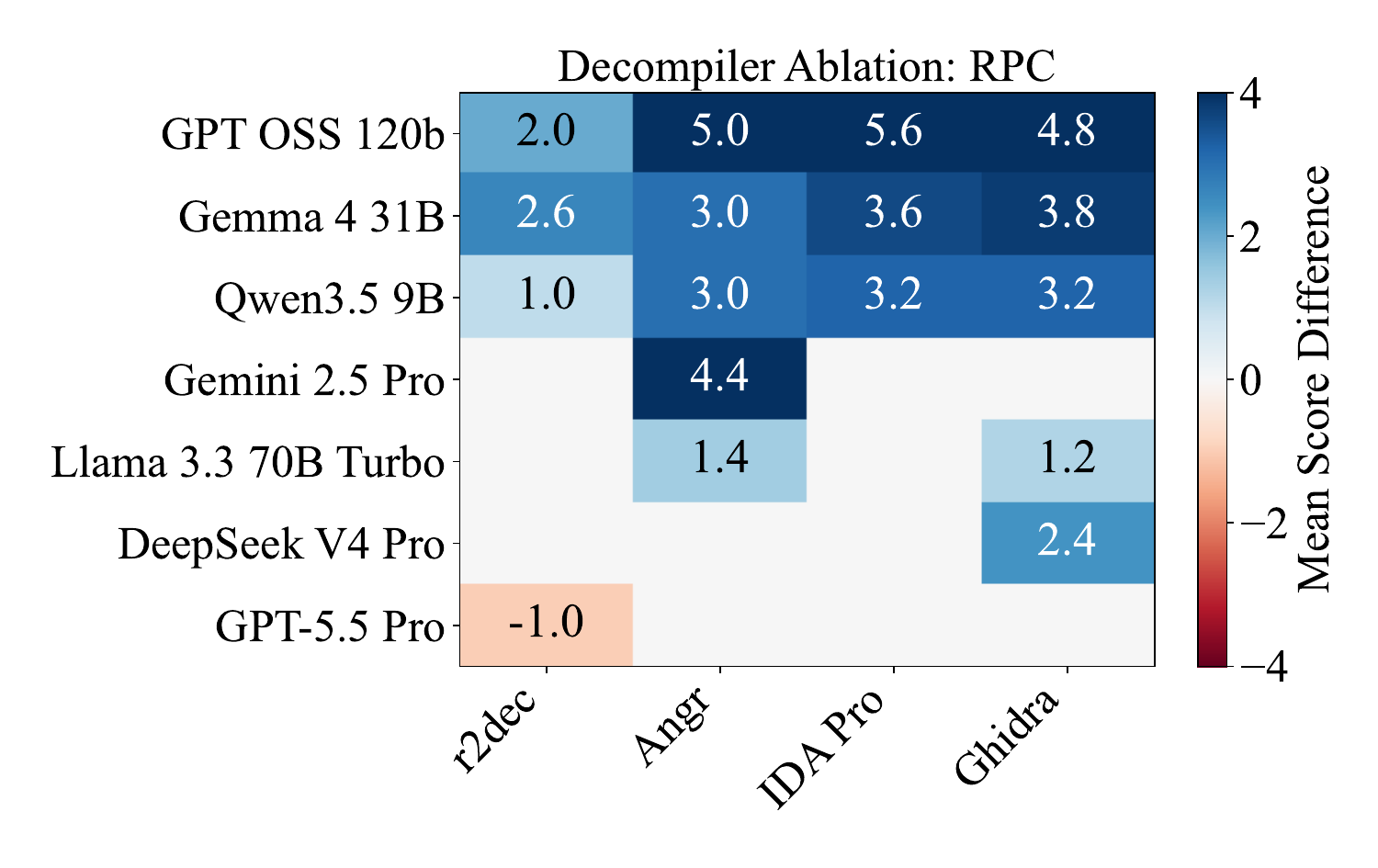}%
    \hfill
    \includegraphics[width=0.48\textwidth]{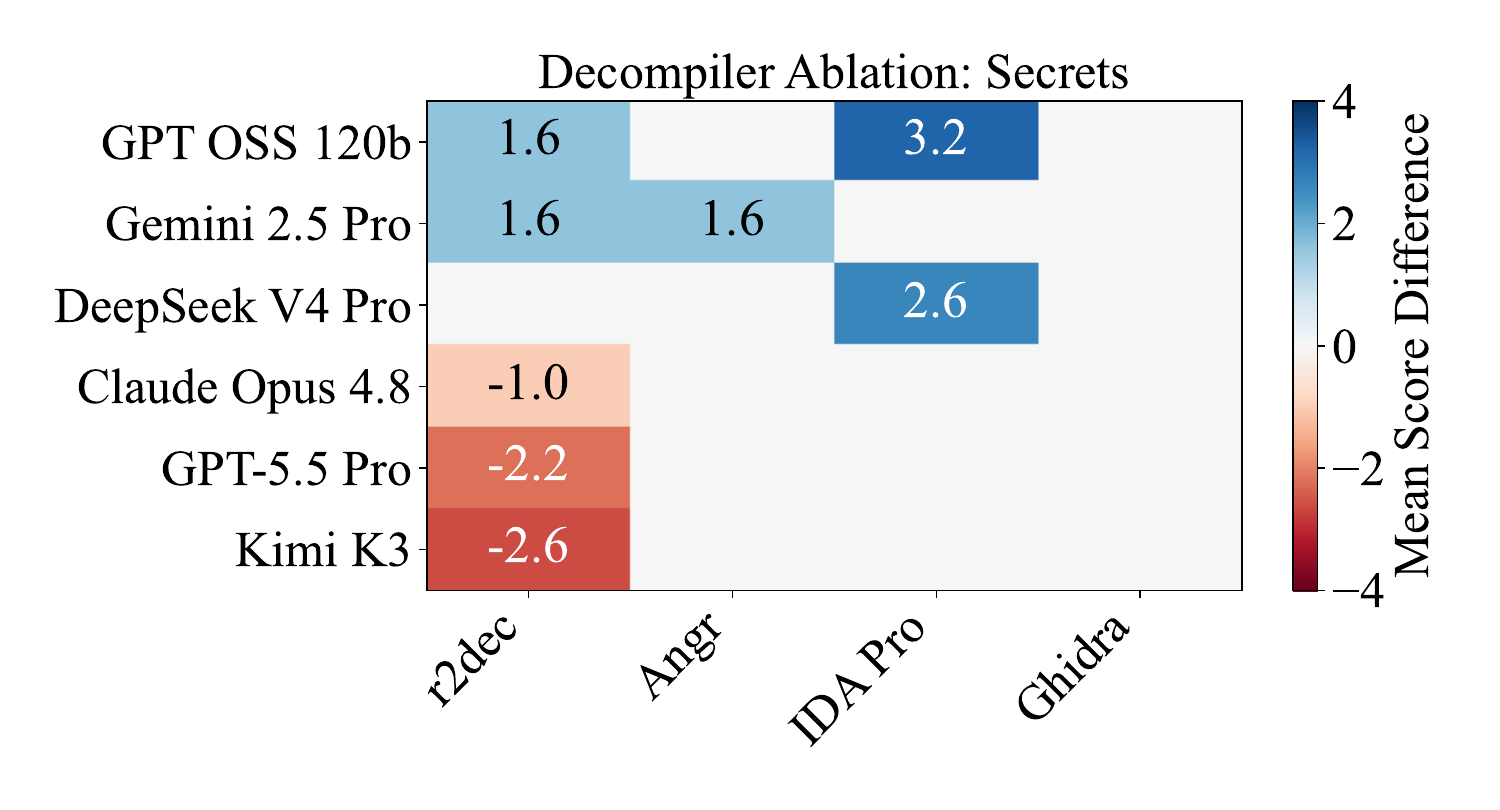}
    \caption{Descriptive per-model understanding-score shifts for full decompiler inputs without MCP relative to baseline MCP runs (\texttt{RPC} left, \texttt{Secrets} right).}
    \label{fig:decompiler_heatmap}
\end{figure}

Figure~\ref{fig:decompiler_heatmap} visualizes the per-model decompiler shifts. Fifteen of the 90 individual static no-MCP contrasts remain significant after BH correction, all with $|\delta|=1$ (i.e., every ablation run outperformed every baseline run for that model) and adjusted $q=0.0476$.
The strongest positive model-level shifts are concentrated in weaker open-weight systems on \texttt{RPC}. Notably, some agents perform \emph{worse} in the no-MCP condition: after correction, Kimi K3 and GPT-5.5 Pro regress under r2dec on \texttt{Secrets}, and Kimi K3 regresses under IDA disassembly on \texttt{Secrets}, suggesting that these strong models benefit from the interactive MCP workflow and/or different code representations on that challenge.

\myparagraph{Human-reference analysis.}
We do not treat $p>0.05$ from a difference test as evidence that an ablated configuration is comparable to humans. Instead, a configuration qualifies as reaching the human reference only if it meets both the same one-point non-inferiority (NI) criterion as RQ1 against the rate-selected top-human cohort ($q_{\mathrm{NI}}$) and a corrected improvement over its own baseline ($q_{\mathrm{shift}}$), both after BH correction.
Only two configurations meet both conditions, both on \texttt{RPC} and using Gemma 4: with full IDA decompilation improves from $3.2$ to $6.8$ ($q_{\mathrm{shift}}=0.0476$) and is non-inferior to the top-human reference ($q_{\mathrm{NI}}=0.00582$); with full Ghidra decompilation it improves from $3.2$ to $7.0$ ($q_{\mathrm{shift}}=0.0476$) and is likewise non-inferior ($q_{\mathrm{NI}}=0.000202$).
No \texttt{Secrets} configuration both significantly improves over baseline and establishes non-inferiority after correction.

\begin{takeawaybox}
    \textbf{RQ2: What Limits Agent Reverse Engineering Performance?}\\[2pt]
    Semantically richer source responses yield a modest aggregate benefit, whereas direct full-decompiler inputs yield substantially larger gains, especially for open-weight agents. These gains are not reproduced by disassembly-only inputs and are not uniform across challenges, models, or representations. The results point to a combination of representation quality, search burden, and tool-mediated workflow demands rather than to any single factor. This suggests that future evaluations of SRE agents should consider model capability and interface design separately, since a weak workflow can hide capabilities that become visible under a better representation.
\end{takeawaybox}

\section{RQ3: How Do Agents' and Humans' Strategies Differ?}
\label{sec:rq3}

Several high-performing agents reach the human reference level, but similar outcomes do not necessarily imply similar analytical processes.
RQ3, therefore, asks how successful agent workflows differ from recorded human reverse-engineering behavior on the same binaries.

In this context, previous work~\cite{shao2026ctfjudge} has evaluated offensive-security agent trajectories against human-crafted step-level references.
Our focus is instead a matched human--agent comparison in which both groups analyze the same binaries, and their behavior can be mapped onto common reverse-engineering abstractions.

\subsection{Macro-Strategies Comparison}
\label{sec:rq3.a}

\noindent \textbf{Methods.} At the macro level, we map both human and agent traces into five categories. Agent traces consist of MCP tool calls logged during each run; human traces consist of manual IDA Pro interactions recorded during the prior study~\cite{basque2026decompiling}. Despite the difference in interaction medium, both expose the same broad analytical activities:
\begin{itemize}
    \item \textit{Recon}: entry-point and binary-overview discovery;
    \item \textit{Data}: inspection of variables, memory, and structures;
    \item \textit{Relationship}: cross-reference and call-graph traversal;
    \item \textit{Read}: direct reading of decompiled or disassembled code;
    \item \textit{Edit}: renaming and annotation.
\end{itemize}
Our categories refine the three-phase model of Votipka et al.~\cite{votipka2020observational}: Recon corresponds to overview; Data and Relationship decompose sub-component scanning; Read captures focused code inspection; and Edit makes state externalization explicit.

For the descriptive macro comparison, we use high-performing human and agent traces to highlight the clearest contrast in strategies. Each session is represented as a sequence over the five categories, from which we compute a first-order Markov transition matrix that shows, for each category, the probability of transitioning to the next. We report $\Delta=P_{\mathrm{agent}}-P_{\mathrm{human}}$, where positive entries indicate transitions that are more common in agent traces, and negative entries indicate transitions more common in human traces. This comparison is \emph{descriptive only}: because we deliberately selected the best-performing traces, any patterns we observe might partly reflect the selection rather than a general behavioral law. We therefore use it only to generate hypotheses about differences in navigation; the formal statistical claims below are based on session-level comparisons that do not condition on the outcome.

\begin{figure}[tb]
    \centering
    \includegraphics[width=0.5\textwidth]{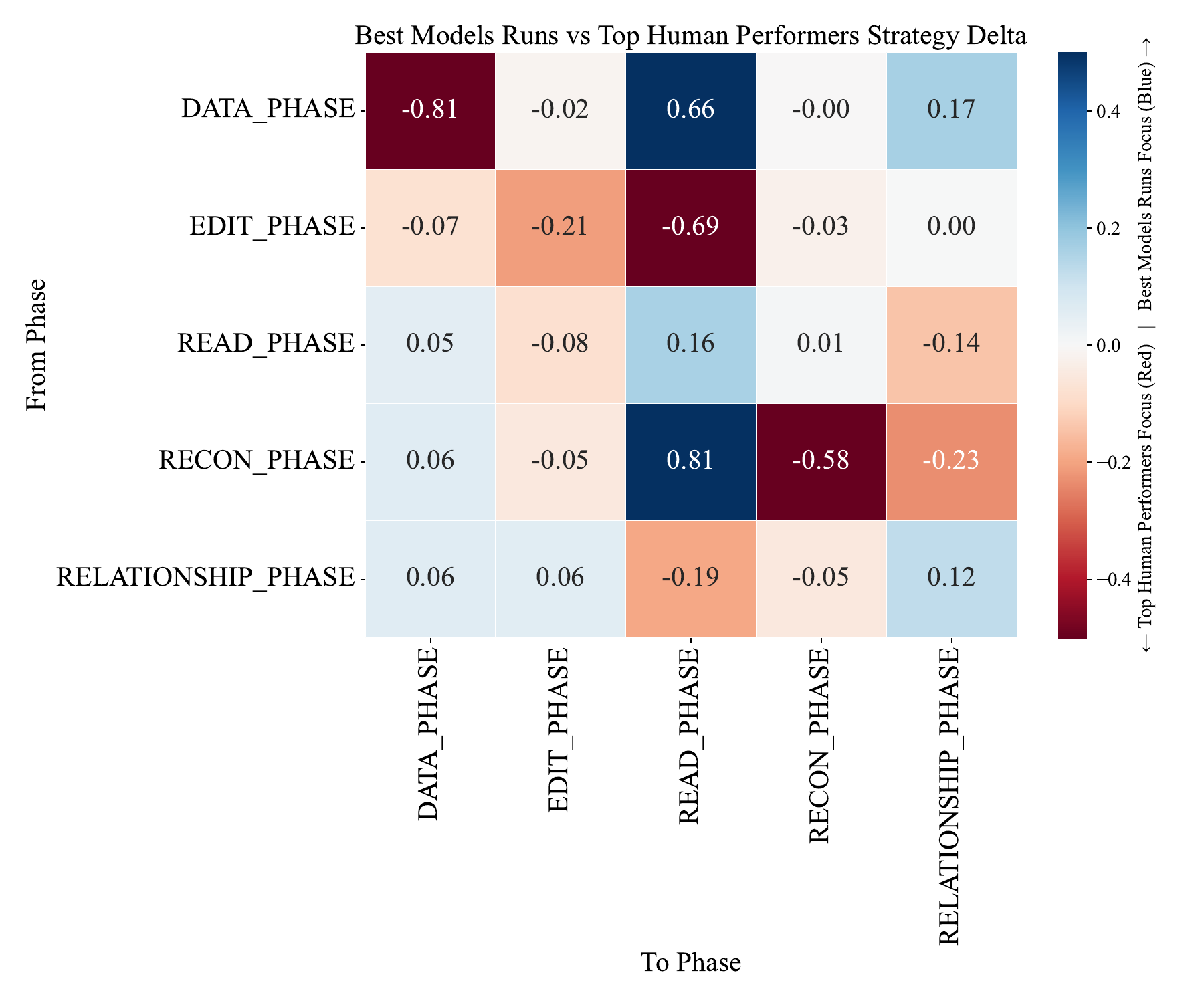}
    \caption{Delta matrix comparing high-performing agent traces with top human traces. Positive values are agent-skewed, and negative values are human-skewed.}
    \label{fig:delta_best_runs_vs_top_humans}
\end{figure}

\noindent \textbf{Results.} Figure~\ref{fig:delta_best_runs_vs_top_humans} shows that high-performing agent traces are more read-centric, whereas human traces allocate more activity to iterative editing and repeated structural exploration.

The strongest agent-skewed transition is \textit{Recon}$\rightarrow$\textit{Read} ($\Delta{=}\DeltaBestRunsVsHumansReconRead$), accompanied by increases in \textit{Data}$\rightarrow$\textit{Read} ($\Delta{=}\DeltaBestRunsVsHumansDataRead$), \textit{Data}$\rightarrow$\textit{Relationship} ($\Delta{=}\DeltaBestRunsVsHumansDataRel$), and \textit{Read} self-loops ($\Delta{=}\DeltaBestRunsVsHumansReadLoop$). Human-skewed behavior includes \textit{Data} self-loops ($\Delta{=}\DeltaBestRunsVsHumansDataLoop$), \textit{Edit}$\rightarrow$\textit{Read} ($\Delta{=}\DeltaBestRunsVsHumansEditRead$), \textit{Edit} self-loops ($\Delta{=}\DeltaBestRunsVsHumansEditLoop$), and \textit{Recon} self-loops ($\Delta{=}\DeltaBestRunsVsHumansReconLoop$). The \textit{Edit} row deserves separate mention: its $\Delta$ entries sum to $-1$ because the high-performing agent traces contain no \textit{Edit} transitions at all. This is not a limitation of the toolset: the MCP interface exposes renaming, retyping, and commenting tools, yet they were never used. Humans externalize what they learn into the database, whereas agents may instead rely on their conversation context to carry analysis state. This descriptive contrast motivates a more explicit analysis of navigation.

\subsection{Micro-Strategies Comparison}
\label{sec:rq3.b}

\noindent \textbf{Methods.} We construct a call graph for each binary and map both agent and human movements between functions. Function depth is the shortest call-graph distance from \texttt{start}. We classify each movement as:
\begin{itemize}
    \item \textit{Down}: from a function to one of its callees;
    \item \textit{Up}: from a function to one of its callers;
    \item \textit{Lateral}: between functions at the same depth without a direct call-graph edge; or
    \item \textit{Jump}: between functions at different depths without a direct call-graph edge.
\end{itemize}
These categories refine the navigation behaviors identified by Mantovani et al.~\cite{mantovani2022re} by explicitly separating hierarchical traversal from non-structural navigation. We further stratify movements by \emph{visit layer} $k$: within a session, a movement out of function $f$ belongs to layer $k$ if it is the $k$-th time the session leaves $f$. Layer $k{=}1$ thus captures how each function is first left, and deeper layers capture how the analyst behaves when returning to functions already visited.

\noindent \textbf{Results.} The layer-based descriptive analysis already suggests a strong navigation asymmetry. At the first visit ($k{=}1$), agents allocate 81.6\% of movements to jumps, compared with 19.0\% for humans; at deeper layers, agent movement remains jump-heavy.
This jump-heavy pattern suggests a strategy we term \emph{greedy fetching}: instead of following call-graph structure step by step, strong agents typically jump to non-adjacent functions, fetch new code, and continue reasoning over newly retrieved information, revisiting relatively little. Greedy fetching describes \emph{how} information is acquired, not simply how much of the binary is eventually covered. On compact binaries, this can be effective because much of the relevant program can be gathered quickly; as programs grow, however, aggressive retrieval may increase tool overhead and consume the available context before the agent can consolidate what it has seen. We treat this as a hypothesis at this stage: the next subsection, RQ4, examines and measures its scalability.

\subsection{Measuring Greedy Fetching}
\label{sec:greedy-fetching}

\noindent \textbf{Methods.} We formalize \emph{greedy fetching} as a four-dimensional session-level behavioral signature based on the following metrics:

\begin{align}
    J & = \frac{N_{\mathrm{jump}}}{N_{\mathrm{function\ transitions}}},                     &
    C & = \frac{N_{\mathrm{unique\ viewed\ functions}}}{N_{\mathrm{callgraph\ functions}}},   \\
    Q & = \frac{N_{\mathrm{code\ read\ events}}}{N_{\mathrm{analysis\ events}}},            &
    R & = \frac{N_{\mathrm{repeated\ function\ views}}}{N_{\mathrm{function\ views}}}.
\end{align}
Here, $J$ is the non-call-graph jump rate, $C$ is the function coverage, $Q$ is the read-event share, and $R$ is the revisit rate. $J$, $C$, and $R$ are computed over the same recovered call graph for each binary, so they are defined identically for humans and agents; $Q$ instead relies on mapping events to the paper's Read category (MCP operations for agents, manual IDA events for humans).

To see how and where greedy fetching manifests, we measure these four metrics on the strong-performing cohorts established in RQ1: the top-performing human participants and all five baseline MCP runs from the three non-inferior models (GPT-5.5 Pro, Claude Opus 4.8, Kimi K3).
Because the 15 agent runs (three models $\times$ five runs for each challenge) share underlying models, they are not truly independent. To avoid artificially inflating statistical significance, we average the five runs within each model to yield three independent model-level values, which we compare against the ten human participants. For each metric $M \in \{J, C, Q, R\}$ and each challenge, we estimate the equal-weight agent mean minus the human mean, $\hat{\Delta}_M$.

We quantify the precision of this estimate with a 95\% CI from a \emph{hierarchical bootstrap}: a two-stage resampling in which some agent models are included and some of their five runs contribute, reflecting both model and run variability, with human participants resampled independently.

We assess statistical significance with an exact permutation test: pooling all 13 independent clusters (three agent-model means and ten human participants), we calculate how often all $\binom{13}{3}=286$ possible assignments to the ``agent'' group produce a difference at least as large as observed, then apply BH correction across the eight planned comparisons. Table~\ref{tab:greedy_metrics} additionally reports session-level medians and inter-quartile ranges.

\begin{table*}[tb]
    \centering
    \footnotesize
    \caption{Greedy-fetching metrics: strong-agent cohort (GPT-5.5 Pro, Claude Opus 4.8, Kimi K3) vs.\ top-performing humans. Columns report session-level median [IQR]; $\Delta$ is the equal-weight agent-mean minus human-mean (percentage points, 95\% hierarchical-bootstrap CI); $q$ is BH-adjusted.}
    \label{tab:greedy_metrics}
    \begin{adjustbox}{max width=\textwidth}
    \renewcommand{\arraystretch}{0.92}
    \begin{tabular}{llcccc}
        \toprule
        Challenge & Metric & Strong agents: median [IQR] & Humans: median [IQR] & $\Delta$ pp [95\% CI] & $q$ \\
        \midrule
        \texttt{RPC} & Jump rate $J$ & 82.1 [69.3, 84.9] & 6.7 [4.5, 11.4] & \textbf{+66.5 [+52.5, +78.4]} & \textbf{0.0070} \\
        \texttt{RPC} & Function coverage $C$ & 96.3 [87.0, 96.3] & 92.6 [85.2, 95.4] & +2.2 [-6.0, +9.8] & 0.7952 \\
        \texttt{RPC} & Read-event share $Q$ & 87.9 [83.1, 88.9] & 72.9 [67.7, 76.2] & +18.5 [+9.0, +29.4] & 0.0727 \\
        \texttt{RPC} & Revisit rate $R$ & 8.0 [4.2, 10.3] & 85.6 [84.1, 87.1] & \textbf{-76.3 [-80.1, -72.2]} & \textbf{0.0070} \\
        \midrule
        \texttt{Secrets} & Jump rate $J$ & 73.9 [58.7, 87.5] & 31.4 [27.1, 42.8] & \textbf{+37.9 [+18.0, +55.4]} & \textbf{0.0070} \\
        \texttt{Secrets} & Function coverage $C$ & 88.2 [82.4, 94.1] & 94.1 [88.2, 98.5] & -0.4 [-12.2, +13.9] & 0.9895 \\
        \texttt{Secrets} & Read-event share $Q$ & 76.5 [71.8, 79.2] & 68.9 [37.0, 76.1] & +17.7 [+2.5, +34.4] & 0.3730 \\
        \texttt{Secrets} & Revisit rate $R$ & 5.9 [0.0, 24.6] & 76.8 [73.6, 81.1] & \textbf{-63.1 [-76.1, -45.7]} & \textbf{0.0070} \\
        \bottomrule
    \end{tabular}
    \end{adjustbox}
\end{table*}

\noindent \textbf{Results.}
The strongest human--agent differences are in \emph{navigation} and \emph{revisitation}, not in how much of these compact binaries is eventually inspected.

On \texttt{RPC}, the strong-agent cohort has a median jump rate of 82.1\% versus 6.7\% for humans, an equal-weight difference of $+66.5$ percentage points. The 95\% CI for this difference is $[+52.5,+78.4]$ (well above zero), confirming agents jump far more often than humans ($q=0.0070$). Instead, the agent revisit rate is 8.0\% versus 85.6\% for humans, a $-76.3$ point difference (95\% CI $[-80.1,-72.2]$; $q=0.0070$), where the entirely negative interval indicates that agents revisit far less.

The pattern repeats on \texttt{Secrets}: jump rate is 73.9\% versus 31.4\% ($+37.9$ points, 95\% CI $[+18.0,+55.4]$, $q=0.0070$), while revisit rate is 5.9\% versus 76.8\% ($-63.1$ points, 95\% CI $[-76.1,-45.7]$, $q=0.0070$).

In contrast, function coverage does \emph{not} differ significantly on either task (\texttt{RPC}: $q=0.795$; \texttt{Secrets}: $q=0.990$), and read-event share differences are not significant after correction (\texttt{RPC}: $q=0.073$; \texttt{Secrets}: $q=0.373$), even though it is higher for agents in the raw numbers.
We obtain the same overall pattern when we include all unassisted human sessions, not just the top ten performers, suggesting the difference reflects a broader contrast between human and agent analysis behavior rather than an artifact of the top-performing cohort selection.

\myparagraph{Greedy fetching depends on the analysis environment.}
We next test whether agents use the same greedy-fetching strategy when the code representation changes.
The source-code replacement experiment (Section~\ref{subsec:rq2-source}) keeps the models and MCP interaction loop unchanged while returning semantically richer original source for code-reading calls. For the same three strong models, all greedy-fetching metrics shift when the code quality improves. Function coverage falls from 89.0\% to 65.0\% on \texttt{RPC} and from 73.1\% to 33.3\% on \texttt{Secrets}; read share falls from 81.1\% to 70.2\% and from 70.5\% to 58.3\%, respectively. Jump rate also falls from 77.6\% to 65.9\% on \texttt{RPC} and from 79.3\% to 73.9\% on \texttt{Secrets}. In other words, when each tool call returns higher-quality code, the agents need to fetch less of it.

Crucially, this reduced fetching does not hurt understanding: the equal-weight score difference between source-code and baseline runs is $+0.267$ on \texttt{RPC} (90\% CI $[0.006,0.528]$) and $+0.067$ on \texttt{Secrets} (90\% CI $[-0.318,0.451]$), both falling entirely within the prespecified $[-1,+1]$ practical-equivalence region: the models understand the binary just as well despite fetching less code.
To test whether these behavioral shifts are statistically significant, we apply the same model-stratified permutation tests used in the human--agent comparison above. After BH correction, function coverage and read share both show significant reductions on both challenges, confirming that agents genuinely fetch less code when given better representations. The jump-rate reduction is significant on \texttt{RPC} but not on \texttt{Secrets}, suggesting that the navigation change is challenge-dependent.

\begin{takeawaybox}
    \textbf{RQ3: How Do Agents' and Humans' Strategies Differ?}\\[2pt]
    Strong agents do not reverse engineer like humans. They achieve similarly broad coverage by jumping aggressively across the binary and rarely revisiting previously inspected code, a strategy we call \emph{greedy fetching}. This behavior is not fixed: richer code representations reduce the need for broad retrieval without reducing understanding. For future SRE agents, the implication is clear: progress may come not from fetching more code, but from better navigation, selective retrieval, and persistent analysis state that helps agents retain and refine what they have already learned.
\end{takeawaybox}

\section{RQ4: How Does Program Size Affect The Feasibility Of Full-Context Analysis?}
\label{sec:rq4}

Greedy fetching depends on a critical assumption: that enough of the binary can be retained within the agent's available context.
RQ4 asks how often complete decompiled or disassembled representations fit within current model context windows when moving beyond the two compact benchmark binaries.
We evaluate this scalability question directly across a large sample of real-world software.

\noindent \textbf{Methods.} We construct four groups of real-world binaries (Windows/Linux and system/application).
To avoid choosing an arbitrary application workload, we ground application selection in public popularity rankings. On Windows, we randomly select and install 50 applications from the Microsoft Store's \emph{Most popular apps for PC} listing available at the time of collection~\cite {microsoft-popular}. On Linux, we follow the same approach on a clean Ubuntu 24.04 system, randomly selecting and installing 50 popular application packages using the Debian Popularity Contest (PopCon) as the popularity source~\cite{debian-popcon}.
On Windows, we distinguish system and application binaries based on standard directory locations. On Linux, we isolate application binaries by hashing directories before and after installing the selected software, removing any binaries already present in the baseline. Following state-of-the-art practice~\cite{dambra2023decoding}, we use Detect-It-Easy~\cite{die} to filter framework and packer code (e.g., \texttt{.NET}, \texttt{PyInstaller}, \texttt{Electron}) that hides the real app's code. 
We remove duplicates by considering their SHA-256 values, randomly select 1,000 binaries per group, and extract complete decompiled and disassembled representations using IDA Pro.
We then tokenize each representation using the appropriate tokenizer for each evaluated model and compute context-window saturation as the ratio of the token count to the model's context window size. A saturation of $100\%$ or higher indicates that the representation fills the entire context window.

\noindent \textbf{Results.} We measure, for each model, the proportion of binaries that saturate the context window (excluding samples for which IDA failed to analyze); the full per-model, per-OS, per-group breakdown is in our repository (\Cref{sec:data-availability}).

Three patterns are consistent: first, decompiled representations are substantially smaller than disassembled representations.
Statistical analysis indicates that disassembly is significantly larger than decompiled code ($p = 1 \times 10^{-5}$, effect size $es = 0.87$, classified as a large effect), approximately twice as large, confirming that decompiled representations use the available context more efficiently and are therefore more compatible with greedy fetching than assembly-level views.

Second, Linux binaries exhibit lower saturation rates than Windows binaries: decompiled code size is significantly larger on Windows ($p = 1.1 \times 10^{-163}$, $es = 0.55$, large effect size), which explains the higher saturation rates and makes a randomly selected Windows binary substantially more likely to exceed the practical operating range of greedy fetching.

Third, although application binaries tend to have higher saturation rates than system binaries, the practical difference is small despite being statistically significant (mean difference of $7.96$ percentage points, $p=2.57 \times 10^{-15}$, $|es|=0.28$, small effect size).

Taken together, these patterns show that greedy fetching through complete binary retrieval is only conditionally feasible in practice.
Even for models with context windows of at least $1M$ tokens (e.g., GPT-5.5 Pro, Kimi K3), the share of binaries that saturate the window on decompiled code ranges from 1.9\% for Linux system binaries to 39.1\% for Windows applications, so complete decompiled retrieval occurs in 60.9-98.1\% of cases depending on the binary group. 
For Single-GPU models, the share of saturating binaries reaches 66.6\% on Windows applications and falls to 20.5\% on Linux applications.

The practical implication is that context saturation is a scalability boundary rather than the primary explanation for failures on the compact benchmark tasks. An agent that retrieves large amounts of decompiled code can fit the entire binary into context in many real-world cases, but in the remaining cases, more selective retrieval and state management are required than greedy fetching alone provides.

\begin{takeawaybox}
\textbf{RQ4: How Does Program Size Affect The Feasibility Of Full-Context Analysis?}\\[2pt]
    Full-context analysis works surprisingly often, but it does not scale indefinitely. Large context windows can hold the complete decompilation of many real-world binaries, yet larger applications, especially on Windows, still exceed this limit, and disassembly further exacerbates it. This marks a practical boundary for greedy fetching: future SRE agents cannot rely on simply loading more code. They will need to decide \emph{what} to retrieve, \emph{what} to keep, and \emph{what} to revisit, combining selective retrieval with persistent analysis state as programs grow.
\end{takeawaybox}
\section{Takeaways}
\label{sec:discussion}

Taken together, our results do not show that autonomous SRE has been solved. They do, however, fit a broader trend reported in prior work: autonomous agents are making inroads into a domain that traditionally depended on skilled human reverse engineers~\cite{chen2025clearagent,liu2026binrex,nishizaka2026llmresistant,radey2026agenticre}. 

\myparagraph{1. Human-competitive SRE is possible, but not yet general.}
RQ1 shows that, on these two compact static-analysis SRE tasks, human-level performance is strongly model-dependent. GPT-5.5 Pro, Claude Opus 4.8, and Kimi K3 satisfy the non-inferiority criterion on both challenges, but the large differences among agents make it clear this capability should not yet be generalized to smaller, workstation-scale LLMs. While Kimi K3 shows open-weight models can achieve this, its size still requires accelerator infrastructure; bringing this capability to ordinary local hardware remains an important future direction for analyzing sensitive software.

\myparagraph{2. Agent capability depends on tool and analysis fidelity.}
RQ2 shows that model quality alone does not explain performance: weaker agents either never establish a productive tool-use loop or retrieve information but fail to turn it into understanding, so autonomous SRE depends on both reasoning capability and the ability to operate the analysis workflow.
Our ablations clarify this: providing the complete decompiled program directly yields substantially larger improvements than returning the original source code via MCP, particularly for open-weight models. Removing MCP alone is not sufficient; representation quality, search burden, tool orchestration, and reasoning ability all contribute. Future agents must therefore treat the interface as an integral part of the system.

\myparagraph{3. Strong agents reach similar coverage through a very different strategy.}
Humans and strong agents eventually inspect similarly broad portions of the binaries, but they reach that coverage in very different ways: agents exhibit substantially more non-call-graph jumping and substantially less revisitation, a behavior we call \emph{greedy fetching}, rather than repeatedly traversing structure and returning to inspected code; agents aggressively move toward new functions and reason over what is accumulated in context.
While prior work has found that sequential exploration is characteristic of human novices~\cite{mantovani2022re}, agents bypass structured exploration entirely: when sufficient code fits within the context, broad exposure can substitute for the structured memory-management strategies humans need.

\myparagraph{4. Greedy fetching is an adaptation, not a fixed property of agents.}
The source-code intervention reveals greedy-fetching changes when the information environment changes: when strong models receive semantically richer source responses via MCP, function coverage and read activity decrease significantly, while understanding remains the same. Agents do not fetch aggressively out of an immutable preference for reading code; rather, greedy fetching is useful when information is fragmented, and retrieval is cheap, so richer information means agents need less of it. Interfaces that expose better abstractions may reduce unnecessary exploration.

\myparagraph{5. Context capacity changes the agent's SRE strategy.}
Across real-world binaries, complete decompiled representations often fit within large context windows, but saturation becomes substantially more common for large Windows apps. This bounds the greedy fetching strategy: when programs exceed the context budget, agents must decide what information to enter into context, discard, or keep for later reasoning. Larger context windows can move this boundary, but do not remove it. Next-generation SRE systems will therefore need mechanisms that current agents use weakly: \emph{selective retrieval}, \emph{persistent analysis state}, and \emph{evidence-driven revisitation}.

\section{Threats to Validity}
\label{sec:threats-to-validity}

\myparagraph{Task scope and external validity.}
We measure SRE performance using two compact, static-analysis CTF-style challenges. This provides a controlled, reproducible setting, but does not cover all of real-world SRE. Our results do not imply that agents perform robustly on all security tasks. However, this restricted setting enables our matched human--agent comparison using the same understanding rubric. While broader benchmarks exist~\cite{liu2026binrex,shao2026ctfjudge,tang2026assistance}, they lack the matched human writeups and fine-grained IDA traces required for our behavioral analysis. RQ4 partially addresses this by evaluating whether complete representations of real-world binaries fit in model contexts, though it measures scalability rather than reasoning.
Furthermore, the evaluation interface itself is static-only, with no debugger or dynamic/runtime execution path (Section~\ref{setup:architecture}), because it intentionally matches the static IDA-based setting used in the human reference study. This does not imply that dynamic analysis is unimportant for real-world SRE; extending this comparison to dynamic tooling is left to future work.

\myparagraph{Training-data contamination.}
Because the challenge binaries and their source code were released publicly before some models' training cutoffs, we cannot rule out the possibility that models were exposed to the underlying implementation during training. The structured 7-point scoring rubrics, ground-truth solution writeups, and fine-grained human interaction logs, by contrast, were never made public, which reduces the risk that models memorized the specific evaluation criteria or reference solutions. This does not, however, eliminate the possibility of prior exposure to the source code itself, and we treat training-data contamination as an unresolved threat to validity rather than a risk we can rule out.

\myparagraph{Human reference and non-inferiority margin.}
RQ1 defines the strongest human reference using the understanding-rate measure~\cite{basque2026decompiling}, which determines \emph{who enters the reference cohort}, while the RQ1 outcome itself is the raw understanding score. Our $\Delta=1$ margin represents a single understanding concept, as even strong humans sometimes miss a rubric item, so non-inferiority indicates practical comparability under this explicit tolerance, not equivalence to all professional reverse engineers.
Relatedly, the RQ2 ablations change the overall analysis environment rather than isolating single causal factors: providing the original source restores names, types, and structure (an upper-bound representation experiment rather than a pure decompiler test), while the no-MCP experiment removes tool-selection costs, search effort, and interaction failures while changing context organization. We therefore interpret these ablations as environmental changes rather than clean causal decompositions.

\myparagraph{Comparing human and agent behavior.}
RQ3 compares manual IDA interactions with agent MCP interactions; the raw events are not directly equivalent, so we address this by mapping behaviors to shared abstractions ($J$, $C$, $R$) based on the recovered call graph. Read share ($Q$) remains the least directly comparable due to differences in granularity, so we do not rely on it alone to claim strategy differences. Our behavioral reference is the ten strongest unassisted participants per challenge~\cite{basque2026decompiling}, who should not be conflated with the general professional population; to ensure statistical validity, we perform inference at the model-cluster level to avoid pseudo-replication, use hierarchical bootstrap intervals, and apply BH correction across the eight planned comparisons.
On interpretation: strong agents jump between structurally unrelated functions more often than humans and revisit functions less often, without achieving significantly broader function coverage, so greedy fetching is an alternative method of acquiring information; RQ3 shows this behavior diminishes with richer program information, indicating it is an environmental adaptation rather than an intrinsic agentic property.

\myparagraph{Model reproducibility.}
Evaluating closed-frontier models affects long-term reproducibility as providers update their systems; we mitigate this by reporting experimental configurations and evaluating open-weight models, but exact reproduction of closed-model behavior may become impossible.

\section{Conclusion}
\label{sec:conclusion}

Autonomous SRE is no longer only a question of whether an agent is capable enough. Our results show that, under this controlled two-task setting, some agents can already reach human-competitive understanding on compact, static-analysis SRE tasks, and do so much faster, but their success depends strongly on the surrounding analysis environment.

We identify \emph{greedy fetching} as a distinct agent strategy: strong agents jump aggressively across a binary, retrieve new information quickly, and revisit far less than human analysts. This behavior changes with richer representations and eventually runs into context limits as programs grow, so future progress will require better interfaces, selective retrieval, persistent analysis state, and more deliberate revisitation, not simply larger context windows.

If humans and machines can reach the same understanding through fundamentally different paths, then the question is no longer only whether machines can reverse engineer, but what \emph{understanding} means when the path to it is no longer uniquely human.

\section{Ethical Considerations}
\label{sec:ethics}

Our study constitutes human subject research (HSR), so we follow our institution's HSR guidelines and received approval from the Institutional Review Board (IRB).
In accordance with our IRB's requirements and our own ethical considerations regarding the proper handling of our participants, all participants were fully informed of the study's purpose, data handling procedures, and their right to withdraw at any time without penalty.
They provided explicit consent for recording and data storage, and all recordings were stored on access-controlled servers in accordance with the best security practices.

\section{Data Availability}
\label{sec:data-availability}

To support transparency and reproducibility, we release an anonymized repository~\footnote{\url{https://anonymous.4open.science/r/humans_vs_machines/}} containing all code, scripts, and agent results required to reproduce our experiments and analyses. The repository includes the evaluation pipeline, data processing and statistical analysis code, and instructions for running the full workflow. It also provides the prompts, configurations, and experimental settings used for all agents. The repository also contains additional details about the statistical methodology, along with the prompts used, the tool list, and additional figures and tables~\footnote{\url{https://anonymous.4open.science/r/humans_vs_machines/supplementary/README.md}}.

\bibliographystyle{ACM-Reference-Format}
\bibliography{biblio}

\end{document}